\documentclass[letterpaper, 10 pt, journal, twoside]{ieeetran}  % Comment this line outiate\frac{\ie}{•}~
\pdfoutput=1

\usepackage{amsmath}
\usepackage{amssymb}
\usepackage{tikz}
\usetikzlibrary{calc, shapes, backgrounds}
\usepackage{epsfig}
\usepackage{verbatim}
\usepackage{graphicx}
\usepackage{subcaption}
\usepackage{algorithm}
\usepackage{algpseudocode}
\usepackage{algorithmicx}
\newtheorem{definition}{Definition}
\newtheorem{theorem}{Theorem}
\newtheorem{lemma}{Lemma}

\newtheorem{proposition}{Proposition}

\usepackage[nodisplayskipstretch]{setspace}
\newcommand{\ie}{\emph{i.e., }}

\def\1{{\bf 1}_N}
\def\0{{\bf 0}}
\def\A{\mathcal{A}}

\def\E{\mathcal{E}}
 
 \def\V{\mathcal{V}}
 \def\X{\mathcal{X}}
\title{Marketing resource allocation in duopolies over social networks\thanks{This work has been funded by the CNRS PEPS project YPSOC}}

\author{Vineeth S. Varma, Irinel-Constantin Mor\u{a}rescu, Samson Lasaulce, and Samuel Martin \thanks{V. S. Varma, I-C. Mor\u{a}rescu and S. Martin are with the Universit\'e de Lorraine, CNRS, CRAN, F-54000 Nancy, France, {\small \tt \{vineeth.satheeskumar-varma\}@univ-lorraine.fr.}}
\thanks{S. Lasaulce is with the Laboratoire des Signaux et Systemes (L2S, CNRS-CentraleSupelec-Univ. Paris Sud), Gif-sur-Yvette, France.} }% 

\begin{document}

\maketitle
\thispagestyle{empty}

\begin{abstract} One of the key features of this paper is that the agents' opinion of a social network is assumed to be not only influenced by the other agents but also by two marketers in competition. One of our contributions is to propose a pragmatic game-theoretical formulation of the problem and to conduct the complete corresponding equilibrium analysis (existence, uniqueness, dynamic characterization, and determination). Our analysis provides practical insights to know how a marketer should exploit its knowledge about the social network to allocate its marketing or advertising budget among the agents (who are the consumers). By providing relevant definitions for the agent influence power (AIP) and the gain of targeting (GoT), the benefit of using a smart budget allocation policy instead of a uniform one is assessed and operating conditions under which it is potentially high are identified.
\end{abstract}
\section{Introduction}

\IEEEPARstart{D}{uopoly} is a standard scenario in economics, politics, and marketing that considers the competition between two (dominant) players over a market, for example, see \cite{singh1984price}. Illustrative examples of real-life duopolies are Airbus/Boeing in the market of large commercial airplanes, Republican/Democratic parties in the American politics. 

Traditional research on competitive games between marketers assumes a homogeneous population of consumers  \cite{friedman1958game,esmaeili2009game}. Unlike these works, we propose a marketing resource allocation based on the influence power that each individual has over the (physical or digital) social network. Basically, we consider that the advertising is done in two steps: the first is done by the marketer that allocate her resources to sway some individuals/agents on her opinion and the second is done by the agents of the social network who influence each other. Consequently, each marketer has to target appropriate influential agents in the network in order to optimize her revenue. Since the focus of the paper is on the resource allocation of the marketer, the second step is modeled by a simple opinion dynamics model introduced in \cite{DeGroot}. 

There is an obvious gap between the literature of economics and the literature on formal opinion dynamics seen from the control community perspective. On one hand, the literature of economics on the considered problems contains many ideas, concepts, approaches but is not very formal, see \cite{tsang2004consumer,woerndl2008internet} which illustrate this point. On the other hand, the control literature is formal but does not address aspects such as the problem of competition over social networks as we do in the manuscript. Therefore, our approach can be seen as a contribution to bridging this gap and the model we propose can be seen as the first step in this direction. Although some recent studies propose the control of one or few agents (see \cite{Camponigro,Dietrich}) in order to enforce consensus, there are very few that deal with the control of opinion dynamics. Besides these methods of controlling opinion dynamics towards consensus, we also find recent attempts to control the discrete-time dynamics of opinions such that as many agents as possible reach a certain set after a finite number of influences \cite{Hegselmann}. In the literature on \emph{viral marketing}, the idea that members of a social network influence each other’s purchasing decisions have been studied, with the goal being to select the best set of people, such that marketing to this set would maximize the overall profit by propagation of influence through the network \cite{domingos2001mining}. This problem has since received much attention, including both empirical and theoretical results \cite{arthur2009pricing}, but these results often consider a single entity influencing the network.

In this paper, we consider two competing marketers who want to use their marketing budget in order to sway on their side as many individuals of the network as possible. Thus, the natural framework to exploit is that of game theory and a reasonable solution concept (for arguments see \emph{e.g.}, \cite{lasaulce-book-2011}) for analyzing such a competition situation is the Nash equilibrium (NE). In \cite{masucci2014strategic}, the authors consider multiple influential entities competing to control the opinion of consumers under a game theoretical setting. However, this work assumes an undirected graph and a voter model for opinion dynamics resulting in strategies that are independent of the node centrality (\ie agent influence power). On the other hand, the recently published work \cite{varma2017opinion} considers a similar competition with opinion dynamics over a directed graph but with no budget constraints and by considering the average agents' opinion instead of the final one; these two differences change the problem significantly.

The main contributions of this paper are the followings. We introduce and analyze a {\bf new generic model for marketing over social networks}. We conduct a {\bf complete equilibrium analysis (existence, uniqueness, determination)} for the corresponding model. We conduct a {\bf numerical performance analysis} that allows one, in particular, to obtain very useful insights in terms of investment for the marketers.

{\bf Notation.} Let $\mathbb{R}_{\geq 0}:= [0,\infty)$ denote the set of non-negative real numbers. If $f(t)$ is a lower semi-continuous function at $t_0$, we use the notation $f(t^+_0)$ to imply $f(t^+_0):= \lim_{t \to t_0, t>t_0} f(t)$. Since we are concerned with a duopoly in this work, for ease of exposition, we will denote by $-i$ when $i \in \{1,2\}$ is a player index, to refer to the index of the other player, i.e. $-i := 1+i(mod\ 2)$.
%We use $\Pr(\cdot)$ to denote the probability, and $E(\cdot)$ for the expectation of a random variable. 

\section{Problem statement}
\label{sec:model}

We consider a market with Firms $1$ and $2$ that are interested in attracting consumers (referred to as \textit{agents}) to their product. We consider a set of $N$ agents that continuously interact over a fixed social network.  In the sequel we denote by $\V= \{1,2,\dots,N \}$ the set of agents and the social network is represented by a fixed weighted directed graph $(\V, \E, \Omega )$, with $\mathcal{E}$ and $\Omega$ respectively representing the set of edges of the graph and the matrix of corresponding weights. 
To agent $n \in \mathcal{V}$ we assign a normalized scalar opinion $x_n(t) \in (0,1)$ which denotes the opinion in favor of the product or ideas of Firm $1$. The revenue obtained by a firm is assumed to be proportional to its average market share. Thus, for agent $n$ at time $t \in \mathbb{R}_{\geq 0}$, the revenue of Firm $1$ is proportional to $x_n(t)$ and for Firm $2$ the revenue is proportional to $1-x_n(t)$. We use $x(t) = (x_1(t),x_2(t),\dots,x_N(t))^\top$ to denote the state of the network at any time $t$, where $x(t) \in \X_0$ and $\X_0 = (0,1)^N$. We can define $x_{n;i}(t)$ as the opinion of agent $n$ in favor of Firm $i$, with $x_{n;1}(t)=x_n(t)$ and $x_{n;2}(t)=1-x_n(t)$, i.e., the two firms have competing products.

\subsection{External influence model (through marketing campaigns)}

In order to obtain a larger market share, Firm $i$ invests according to the investment or action vector which corresponds to the marketing campaign. Without loss of generality we can consider that the campaign is reduced to a time instant and it has an impulsive instantaneous effect on the opinion of the agents. At the campaign instant, Firm $i$ invests according to the vector $a_i\in \A_i$ where the action space for Firm $i$ is
\begin{equation}
\A_i := \left\{a_i  \in [0,b_i]^N| \sum_{n=1}^N a_{i,n} \leq B_i  \right\}
\label{eq:defAset}
\end{equation}
where $b_i\le B_i \in \mathbb{R}_{\ge0}$ for $i \in \{1,2\}$ represent the maximum influence/discount for one specific agent and the total budget, respectively. The vector $a_i,\ i \in \{1,2\}$ is called the \emph{action} of Firm $i$, with $a_{i,n}$ being the marketing expenditure targeted at agent $n$. The campaign modify the opinions of agents according to a function $\Phi(x_0, a_i,a_{-i}):  \X_0  \times \A_i \times \A_{-i} \to \X_0 $ where $x_0 = (x_{0,1},...,x_{0,N})^{\mathrm{T}} \in \X_0$ is the vector collecting the initial opinions of the agents before the campaign in favour of Firm $i$, i.e. $x_0=x(0)$. %Although the marketing campaign may last for a certain duration in practice, for the sake of exposition of results, we treat this as an instantaneous jump in the opinions. 
Without any loss of generality, we consider that the campaign occurs at $t=0$. In the sequel we consider the function $\Phi(x_{0;i}, a_{i},a_{-i}) = (\phi(x_{0,1;i}, a_{i,1},a_{-i,1}),..., \phi(x_{0,N;i}, a_{i,N},a_{-i,N})  )^{\mathrm{T}}$ with
\begin{equation}
\phi(x_{0,n;i}, a_{i,n},a_{-i,n})= \frac{x_{0,n;i}+ a_{i,n}}{1+a_{i,n}+a_{-i,n}},\ \forall n \in\{1,...,N\}. 
\end{equation}
%where  \textcolor{blue}{$a_{i,n} \in [0,b_i]$}, $i\in\{1,2\}$, represents the marketing resource allocation vector of Firm $i$ towards Agent $n$. 
If $x_n(t)$ is seen as the probability of agent $n$ picking the product of Firm $1$, $\phi(\cdot)$ corresponds to a Bayessian update rule on the opinion as used in \cite{martins2008continuous}. This probability is updated at the campaign instant $t=0$, with $a_{1,n}$ increasing the odds of choosing Firm $1$, and $a_{2,n}$ increasing the odds of choosing Firm $2$. Note that $$\phi(x_{n;i}(0),a_{i,n},a_{-i,n})=1-\phi(x_{n;-i}(0),a_{-i,n},a_{i,n}),$$ and therefore, the opinion's change is symmetric w.r.t the two firms. This results in the campaign opinion's change
\begin{equation}
x_{n;i}(0^+) = \phi(x_{n;i}(0),a_{i,n},a_{-i,n}) 
\label{eq:camdyn}
\end{equation}
for all $n \in \mathcal{V}$. The proposed function also satisfies the following properties. If both firms spend very little on agent $n$, its opinion at $0$ is preserved after the campaign as we have
\begin{equation}
\lim_{a_{i,n} \to 0, a_{-i,n} \to 0} \phi(x_{0,n;i}, a_{i,n}, a_{-i,n}) = x_{0,n;i}.
\end{equation}

Additionally, when the resources allocated to agent $n$ are large by both firms (provided that $b_i$ is large), the final opinion depends only on the ratio of investments and not on the initial opinion.
\begin{equation}
\lim_{a_{i,n} \to \infty,  a_{-i,n} \to \infty, \frac{a_{-i,n}}{a_{i,n} } \to c }
 \phi(x_{0,n;i}, a_{i,n}, a_{-i,n}) = \frac{1}{1+c}
\end{equation} 

\subsection{Internal influence model (within the social network)}

After the campaign, the consumer's opinion is only influenced by the other consumers of the networks. In this work we consider that opinion dynamics in the social network is characterized by a linear model described by the following differential equation: 
\begin{equation}
\dot{x}(t)= -\mathbf{L} x(t)
\label{eq:opdyn}
\end{equation}
where $\mathbf{L} \in \mathbb{R}^{N \times N}$ is the Laplacian matrix associated with the graph $(\mathcal{V}, \mathcal{E}, \Omega )$, whose components are defined as:
\begin{equation}
L_{m,n}=\left\{\begin{array}{c}
\displaystyle\sum_{n=1}^N \Omega_{m,n} \mbox{ if } m=n\\[2mm]
-\Omega_{m,n}  \mbox{ if } m\neq n
\end{array}\right..
\end{equation} 
As a result of the marketing campaigns, we have the following hybrid opinion dynamics model
\begin{equation}
\left\{\begin{array}{llll}
\dot{x}(t)&=& -\mathbf{L} x(t)& \forall t \in \mathbb{R}_{\geq 0} \setminus \{0\}\\
x_n(0^+) &= &\phi(x_{0,n}, a_{1,n},a_{2,n})&  \forall n \in \mathcal{V} 
\end{array}\right..
\label{eq:opdyn}
\end{equation}

\subsection{Revenue model}

In this work, we assume that the profit is based only on the opinion the agents have after some time $T>0$. Other profit models such as an integral over time for the opinion have been studied in \cite{varma2017opinion}. In either case, we observe that the profit can be expressed as a linear combination of the opinions after the campaign. We define the \emph{agent influence power} (AIP) of Agent $n$ as follows.
\begin{definition}
The AIP of Agent $n$ is given by $\rho_n>0$ where
\begin{equation}
\rho= {1}_N^\top \exp(-\mathbf{L}  T) \label{eq:defrho}
\end{equation}
where ${1}_N$ is a column vector of ones.
\end{definition}
The net revenue earned by Firm $i$ can therefore be written as the difference between the profit and the marketing expenses resulting in
\begin{equation}
u_1(x_0,a_{1},a_2) := \gamma_1 \rho x(0^+)  -  \lambda_1 {1}_N^\top  a_{1},
\label{eq:uiv1}
\end{equation}
\begin{equation}
u_2(x_0,a_{1},a_{2}) := \gamma_2 \rho [{1}_N-x(0^+)]  -  \lambda_2  {1}_N^\top  a_{2}.
\label{eq:u2v1}
\end{equation}
where $x_{n}(0^+)= \phi( x_{0,n},a_{1,n},a_{2,n} )$, $\gamma_i \geq 0$ is the revenue generated per consumer for Firm $i$, and $\lambda_i \geq 0$ is the advertising efficiency or pricing factor for Firm $i$.

\subsection{Game-theoretic formulation of the problem}

The ingredients introduced in this section allow us to formulate the problem as a game under strategic form that is, a triplet which is defined as follows: 
\begin{equation}
\mathcal{G} =  \left(\{1,2\}, \{ \A_1,\A_2 \},    \{ u_1,u_2 \} \right),
\end{equation} 
where:
\begin{itemize}
\item $\{1,2\}$ is the set of players (i.e., Firms $1$ and $2$);
\item $\A_i$ defined in \eqref{eq:defAset} is the set of pure actions for Player $i$;
\item $u_i$ as defined per (\ref{eq:uiv1}) \eqref{eq:u2v1} is the utility function for Firm $i$.  %as defined per (\ref{eq:uiv1}) \eqref{eq:u2v1} and which  can be rewritten as
%\begin{equation}
%u_{i}(x_{0;i},a_i,a_{-i}) = \gamma_i \sum_{n=1}^N \rho_n \frac{x_{0,n;i} + a_{i,n}}{1+a_{i,n} + a_{-i,n}} - \lambda_i a_{i,n}.
%\end{equation}
\end{itemize}

Throughout the paper we assume non-aligned utility functions and partial control for the players. In other words, in the game theoretical framework, the meaning of optimality is not clear and there is a need for defining a solution concept for the considered problem (see \emph{e.g}., \cite{lasaulce-book-2011} for further discussion). The solution retained here is the Nash equilibrium (NE) \cite{nash1951non}, as commonly assumed for duopolies. The definition of a pure NE is as follows.

\begin{definition}[Pure NE]\label{def:PNE} A strategy profile $(a_1^\star,a_2^\star) \in \A_1 \times \A_2$ is a pure NE for $\mathcal{G}$ for a given $x_0$ if $\forall i \in \{1,2\}, $
\begin{equation}
\ \forall a_i \in \A_i, \
u_i(x_0, a_i^\star, a_{-i}^\star) \geq u_i(x_0, a_i, a_{-i}^\star).
\end{equation}
\end{definition}

%==========================================
%==========================================
%==========================================
\section{Game-theoretic analysis}
\label{sec:gt-analysis}

This section analyse whether the solution of the problem formulated above exists, is unique and how it can be attained.

\subsection{Nash equilibrium analysis}

We start by showing that the game $\mathcal{G}$ does have a pure and unique NE. First, we provide some auxiliary lemmas that will help us to characterize the utility function.
\begin{lemma} \label{lemma1}
The utility function $ u_i (x_{0;i},a_i,a_{-i})$ is concave w.r.t $a_i$.
\end{lemma} 
\begin{IEEEproof} Recall that $x_{0,n;i}$ stands for the initial opinion of agent $n$ w.r.t. Firm $i$ and that $x_{0,n;i} = 1-x_{0,n;-i}$. We evaluate 
\begin{equation}\begin{array}{llr}
\frac{\partial u_i }{\partial a_{i,n}}& = &  \frac{ \gamma_i \rho_n}{1+a_{i,n} +a_{-i,n}} - \frac{ \gamma_i \rho_n(x_{0,n;i} + a_{i,n} ) }{(1+a_{i,n} +a_{-i,n})^2 } - \lambda_i
\end{array}
\label{eq:firstderui}
\end{equation}
Next, we evaluate
\begin{equation}\begin{array}{lll}
\frac{\partial^2 u_i }{\partial a_{i,n}^2}& = & -\frac{ 2\gamma_i \rho_n}{(1+a_{i,n} +a_{-i,n})^2 } + \frac{2\gamma_i\rho_n(x_{0,n;i} + a_{i,n} ) }{(1+a_{i,n} +a_{-i,n})^3 }\\
&=&- \frac{2\gamma_i \rho_n(1-x_{0,n;i} + a_{-i,n} ) }{(1+a_{i,n} +a_{-i,n})^3 }
\end{array}
\end{equation}
Note that $a_{i,n}, a_{-i,n} \geq 0$ and $0 < x_{0,n;i}< 1$ for all $i \in \{1,2\}$, $n\in \{1,2,\dots,N\}$. Therefore, we have $\frac{\partial^2 u_i }{\partial a_{i,n}^2} <0$ for all $n$. We can also easily see that $\frac{\partial^2 u_i }{\partial a_{i,n} \partial a_{i,m}}=0$ for all $m \neq n$. The Hessian of the utility function is therefore a diagonal matrix with all entries negative. Therefore, $u_i(\cdot)$ is concave w.r.t. $a_i \in [0,b_i]^N$. 
\end{IEEEproof}

Next, consider the weighted sum of utilities for the game $\mathcal{G}$ for a given initial opinion $x_0$ defined as
\begin{equation}
\sigma(a,r): =  r_1 u_1(x_{0},a_1,a_2) +r_2 u_2(x_0,a_1,a_2).
\end{equation}
for some $r_1,r_2 >0$. We look at the pseudo-gradient of $\sigma(a,r)$, used by Rosen in \cite{rosen1965existence}, and defined as
\begin{equation}
%\begin{array}{lll}
g(a,r)= 
 \left(\begin{array}{l}
r_1 \nabla _{a_1} u_1(a), 
r_2 \nabla _{a_2}  u_2(a)
\end{array} \right)^\top.
%=  \left(\begin{array}{l}
%r_1 \frac{\partial u_1(a) }{a_{1,1}}, \dots,r_1 \frac{\partial u_1(a) }{a_{1,N}} \\  
%r_2 \frac{\partial u_2(a) }{a_{2,1}}, \dots ,r_2 \frac{\partial u_2(a) }{a_{2,N}}  \end{array}\right).
%\end{array}
\end{equation}
This allows us to look at the generalized diagonally strict concavity (DSC) condition \cite{rosen1965existence}, which can be exploited to prove the uniqueness of the NE.
\begin{definition} A function $\sigma(a,r)$ is said to satisfy the DSC for a given $r > 0$ if for every distinct pairs of action profiles $a$, $a'$, i.e., with $(a_{1},a_2) \neq (a_1',a_2'), x_0=x_0'$, we have
\begin{equation}
(a-a')^{\mathrm{T}} (g(a,r) - g(a',r)) <0.
\end{equation}
\end{definition}
\begin{lemma}
The function $\sigma(a,r)$ satisfies the DSC property w.r.t $(a_1,a_2)$.
\end{lemma}
\begin{IEEEproof} When the utility functions are twice differentiable, a sufficient condition for DSC (see  \cite[Th.6]{rosen1965existence}) is that
\begin{equation}\label{eq19}
G(a,r) +G(a,r)^{\mathrm{T}}<0
\end{equation}
where $G(a,r)$ is the Jacobian of $g(a,r)$ w.r.t. $(a_1,a_2)$ and \eqref{eq19} means that $G(a,r) +G(a,r)^{\mathrm{T}}$ is negative definite.

We choose $r=\left(\frac{1}{\gamma_1},\frac{1}{\gamma_2}\right)$ and we look at the elements of $G(a,r)$. The diagonal elements are $r_i \frac{\partial^2 u_i }{\partial a_{i,n}^2}$ which are negative as we have already shown in Lemma \ref{lemma1}. The off-diagonal elements are in general given by $r_i \frac{\partial^2 u_i }{\partial a_{i,n} \partial a_{j,m}}$. If $m \neq n$, we notice from (\ref{eq:firstderui}) that $\frac{\partial^2 u_i }{\partial a_{i,n} \partial a_{j,m}} =0$ for any $j\neq i$.\\[1mm] However, if $m=n$, then the off-diagonal terms of $G$ at positions $(n,N+n)$ and $(N+n,n)$ for $1 \leq i \leq N$ are non zero and can be evaluated as
\begin{equation}
 \frac{1}{\gamma_i} \frac{\partial^2 u_i }{\partial a_{i,n} \partial a_{-i,n}}  = \rho_n \frac{-1   -a_{-i,n} + 2x_{0,n;i} + a_{i,n} }{(1+a_{i,n} +a_{-i,n})^3 }
\end{equation}

Note that $$-1 +  2x_{0,n;i} = 2 - 2x_{0,n;-i}-1 = - (-1 + 2x_{0,n;-i}) .$$Therefore, the term at $(n,N+n)$ of $G(a,r)$ is opposite to the term at $(N+n,n)$. As a result, $G(a,r) +G(a,r)^{\mathrm{T}}$ becomes a diagonal matrix with all diagonal entries negative, which is therefore a negative definite matrix.
\end{IEEEproof}
\begin{theorem}
The game $\mathcal{G}$ has a pure and unique NE.
\end{theorem}
\begin{IEEEproof} Notice that the action space $\A_i$ defined as $a_{i,n} \in [0,b_i]$ with $\sum_{n=1}^N a_{i,n} \leq B_i$ is a convex and compact subset of $\mathbb{R}_{ \geq 0}^N$. The utility function is therefore concave w.r.t $a_i \in \A_i$ from Lemma 1, with $\A_i$ being a compact and convex set and is also (jointly) continuous in $a$. This allows us to use the result in \cite[Th.1]{rosen1965existence} and prove that the game $\mathcal{G}$ has at least one pure Nash equilibrium. We can use the result in \cite[Th.2]{rosen1965existence} to prove that the NE is unique since the DSC condition shown in Lemma 2 is met if the NE exists.% Since we have already shown that the NE exists, we use Lemma 2 and conclude that the game $\mathcal{G}$ has a unique pure NE.
\end{IEEEproof}

\subsection{Dynamic characterization of the NE}

We have shown that $\mathcal{G}$ has a unique pure NE. %A central issue to be dealt with is to assess the performance of the two firms at equilibrium. 
To evaluate the performance of the two firms at equilibrium it is necessary to determine the NE. While it is not always possible to express the equilibrium actions, it is possible to fully characterize the equilibrium through a set of dynamic equations, which is given next. The corresponding system of equations can either be seen as a way of numerically determining the NE or as a way of modeling the firms economical behavior. Indeed, we propose to exploit the smooth or continuous time best-response dynamics introduced in \cite{gilboa1991social} and used more recently \emph{e.g.}, in \cite{matsui1992best,hofbauer2006best}. A continuous-time best-response dynamics is a set of differential equations which allows one to generate a trajectory converging to the NE. It is given by
\begin{equation}
\dot{a}_{i,n} = \beta_{i,n}(x_{0;i},a_{-i}) - a_{i,n}  \label{eq:contbrd}
\end{equation}
for all $i \in \{1,2\}$ and $n \in \mathcal{V}$. The quantity $\beta_{i,n}(x_{0;i},a_{-i})$ corresponds to the best-response, which is defined by:
$%\begin{equation}
\beta_i(x_{0;i},a_i): = \arg \max_{a_i} \{u_i(x_{0;i},a_i,a_{-i})  \}.
$ We note that (\ref{eq:contbrd}) should be a differential inclusion since the best-response function is in general a set-valued function. However, the argmax operation returns singleton sets (due to strict concavity). By definition, if $(a_1^\star,a_2^\star)$ is a NE of $\mathcal{G}$ at $x_{0;i}$ then
\begin{equation}
a_1^\star = \beta_1(x_{0;1},a_2^\star)\, , \, a_2^\star = \beta_2(x_{0;1},a_1^\star)
\label{eq:NEBR}
\end{equation}
and can be found by computing the unique point in $\A_1 \times \A_2$ where $\beta_{i,n}(x_{0;i},a_{-i}) - a_{i,n} =0$ holds for all $i,n$. \\
Given an action profile $a_{-i}$, the best-response by Firm $i$ can be evaluated by optimizing $u_i$ w.r.t. $a_{i}$ while respecting the budget constraint. Our next proposition gives a method of evaluating $\beta_i$. Denote $\beta_{i,n}(x_{0;i},a_{-i})$ as the $n$-th component of $\beta_i$. Then, we have
\begin{proposition} The best-response functions are given by
\begin{equation}
\beta_{i,n}(x_{0;i},a_{-i}) = \min\{ b_i, \max\{0,\alpha_{i,n}(x_{0;i},a_{-i})   \} \}  \label{eq:brf}
\end{equation}
where
\begin{equation}
\alpha_{i,n}(x_{0;i},a_{-i}) = \sqrt{\frac{\gamma_i \rho_n (x_{0,n;-i} + a_{-i,n} )}{\mu_{0;i}+ \lambda_i} } - 1 -a_{-i,n} \label{eq:alpha}
\end{equation}
for all $n \in \mathcal{V}$, and $\mu_0 \in \mathbb{R}_{\geq 0}$ is such that
\begin{equation*}\label{eq:mu0eq}
\sum_{n=1}^N \beta _{i,n}(x_{0;i},a_{-i}) \leq B_i, \mu_{0;i} ( \sum_n \beta_{i,n}(x_{0;i},a_{-i}) -B_i)=0 
\end{equation*}  
\end{proposition}
\begin{IEEEproof}
The optimization problem to be solved in order to evaluate $\beta_i$ can be written as
\begin{equation}
\begin{array}{lr}
\text{Maximize}_{a_i} u_i(a_i,a_{-i}) &   \\
\text{subject to } \sum_{n=1}^N a_{i,n} \leq B_i  \>\>\> \&  &\\
0 \leq a_{i,n} \leq b_i & \forall n 
\end{array}\label{op1}
\end{equation}
Problem (\ref{op1}) necessarily has an optimal solution as $u_i$ is continuous and the optimization space is a compact set. Since all the constraint functions are linear, Karush-Kuhn-Tucker (KKT) conditions \cite{boyd2004convex} can be applied. Additionally, the problem being convex, KKT conditions are not only necessary but also sufficient conditions for optimality. For all $n \in \{-N,\dots,N\}$ let $\mu_n \geq 0$ be the KKT multipliers. We use $\mu_n$ for the constraint $a_{i,n} \geq 0$, $\mu_{-n}$ for the constraint $a_{i,-n} \leq b_i$ for all $n \in \mathcal{V}$  and $\mu_0 \geq 0$ as the multiplier for the total budget constraint in (\ref{op1}). Then, the KKT conditions can be written as
\begin{equation*}
\left\{\begin{array}{l}
\begin{array}{ll}
\nabla_{a_i} u_i(a_i^\star)= & \sum_{n=1}^N - \mu_n \nabla_{a_i} a_{i,n}^\star+ \mu_0  \nabla_{a_i} \sum_{n=1}^N  a_{i,n}^\star  \\
 &+ \mu_{-n} \nabla_{a_i} a_{i,n}^\star \geq 0,\\
a_{i,n}^\star \leq b_i, & \sum_{n=1}^N a_{i,n}^\star -B_i \leq 0, \\
\mu_n, \mu_{-n} \geq 0, &\forall n \in \{-N,\dots,N\} ,
\end{array} \\
\begin{array}{ll}
\mu_n (a_{i,n}^\star - b_i)=0 ,\ \mu_{-n} a_{i,n}^\star=0 & \forall n \in \{1,\dots,N\}, \\
\mu_0 (\sum_{n=1}^N a_{i,n}^\star -B_i)=0.  \\
\end{array}
\end{array}\right.
\end{equation*}

In order to evaluate $ \nabla_{a_i} u_i(a_i,a_{-i})$, we have
\begin{equation}
\begin{array}{llr}
\frac{ \partial u_i(a_i,a_{-i})}{\partial a_{i,n}} &= & \gamma_i \rho_n \left( \frac{x_{0,n;i} + a_{i,n}}{1+a_{i,n} + a_{-i,n}} \right)' -\lambda_i \\
&= & \gamma_i\rho_n \frac{x_{0,n;-i} + a_{-i,n} }{ (1+a_{i,n} + a_{-i,n} )^2} - \lambda_i.
\end{array}
\end{equation}
As $x_{0,n;-i}$ and $a_{-i,n}$ are constants, we define $d_{i,n} :=\rho_n (x_{0,n;-i} + a_{-i,n} )$. Thus, the first KKT condition can be now written as
\begin{equation}
\frac{d_{i,n}}{ (1+a_{i,n}^\star + a_{-i,n} )^2} = \lambda_1 +\mu_0 +\mu_{n} - \mu_{-n}
\end{equation}
which must be satisfied for all $n \in \mathcal{V}$. Let $\mathcal{W}_0 \subseteq \mathcal{V}$ be the set of agents for which $a_{i,n}=0$ which leaves $\mu_{-n}  \geq 0$ free in order to satisfy the KKT condition for $n \in \mathcal{W}_0$. Similarly, let $\mathcal{W}_1 \subset \mathcal{V}$ such that $a_{i,n} = b_i$ for all $n \in \mathcal{W}_1$. Finally, let $\mathcal{W}_3 := \mathcal{V} \setminus \mathcal{W}_1 \setminus \mathcal{W}_0$.
Then, we have
\begin{equation}
\beta_{i,n}(x_{0;i},a_{-i}) = \left\{ \begin{array}{llr}
\sqrt{\frac{d_{i,n}}{\mu_0+ \lambda_i} } - 1 -a_{-i,n} & \forall n \in \mathcal{W}_2\\
0 & \forall n \in \mathcal{W}_0 \\
b_i & \forall n \in \mathcal{W}_1
\end{array} \right.
\end{equation}
which must be positive. We can use the final complementary slackness condition to solve for $\mu_0$ and get
\begin{equation}
\mu_{0} = \left( \frac{\sum_{n \in \mathcal{W}_2} \sqrt{d_{i,n}} }{B_i - b_i |\mathcal{W}_1| + |\mathcal{W}_2| + \sum_{n \in \mathcal{W}_2} a_{-i,n} }  \right)^2 -\lambda_i.
\label{eq:wls}
\end{equation}
Therefore $ \mathcal{W}_2$ must be chosen such that $0 \leq \beta_{i,n}(a_{-i})  \leq b_i$ for all $n \in \V$. This gives us (\ref{eq:brf}). % concluding the proof. 
We write $\mu_{0;i}$ in the BR function as this parameter is player dependent. Finally, since the problem is concave as shown in Lemma 1, we know that any point in which the KKT conditions are satisfied must also be the global maximum.
\end{IEEEproof}

Practically, the water level $\mu_{0;i}$ can be found with a lower complexity by first sorting all the agents based on $\frac{1+a_{-i,n}}{\sqrt{d_{i,n}}}$. This is because for a given $a_{-i}$, the agent with a lower value of $\frac{1+a_{-i,n}}{\sqrt{d_{i,n}}}$ will necessarily have a larger $\beta_{i,n}$. This lets us enforce that any agent with $0$ resource allocation to have a lower sorting index than the ones based on \eqref{eq:alpha}, which in turn will have a lower sorting index than the ones with allocation $b_i$.
\subsection{Expressing the NE}
Although we have provided a dynamical characterization of the NE from which a discrete-time iterative algorithm that converges to the NE, we have no insights on how this NE depends on the various key parameters of the problem. In particular, we would like to determine the relationship between the structure of the resource allocation policy at NE, the AIP parameter $\rho_n$, and the initial opinion $x_{0,n;i}$.
The following proposition shows that the amount of budget allocated to Agent $n$ at NE increases proportionally with $\rho_n$, and will decrease if the initial opinion of Agent $n$ is already in favor of Firm $i$. The allocation also depends on budget constraints, the profit multiplier $\gamma_i$, and the advertising efficiency $\lambda_i$. 

\begin{proposition}
For each $n \in \mathcal{V}$, the NE $(a_{1,n}^\star,a_{2,n}^\star)$ is given by
\begin{itemize}
\item $(y,0)$ (or $(0,y)$) if $\exists y \in [0,b_1]$ (or $[0,b_2]$ respectively) such that \eqref{eq:NEBR} is satisfied by one of these pairs,
\item or $(y,b_2)$ (or $(b_1,y)$) if $\exists y \in [0,b_1]$ (or $[0,b_2]$ respectively) such that \eqref{eq:NEBR} is satisfied by one of these pairs,
\item or $(a_{1,n}^\star,a_{2,n}^\star) \in (0,b_1) \times (0,b_2)$ and is given by
\begin{equation}
a_{i,n}^\star = \left(\frac{ k_i}{k_i +k_{-i}}\right)^2 k_{-i}\rho_n - x_{0,n;i},
\end{equation}
where $k_i= \frac{\gamma_i}{\lambda_i+\mu_{0;i}}$ and $\mu_{0;i}$ is a common constant for all $n \in \mathcal{V}$ given by \eqref{eq:mu0eq}.
\end{itemize}
\end{proposition}
\begin{IEEEproof}
We observe from \eqref{eq:NEBR} and \eqref{eq:brf} that if $a_{i,n}^\star$ for some $i \in \{1,2\},n \in \mathcal{V}$ is in the interval $(0,b_i)$, then, it must be equal to \eqref{eq:alpha}. Looking at \eqref{eq:alpha}, we observe that at NE, we have
\begin{equation}
\begin{array}{l}
a_{1,n}^\star = \sqrt{k_1 \rho_n(1-x_{0,n}+a_{2,n}^\star)}-1-a_{2,n}^\star  \\
a_{2,n}^\star = \sqrt{k_2  \rho_n(x_{0,n}+a_{1,n}^\star)}-1-a_{1,n}^\star \\
\Rightarrow k_1 (1-x_{0,n}+a_{2,n}^\star) = k_2 (x_{0,n}+a_{1,n}^\star)\\
\Rightarrow a_{2,n}^\star = \frac{k_2}{k_1} (x_{0,n}+a_{1,n}^\star)-1+x_{0,n}
\end{array}
\end{equation}
for all $n \in \mathcal{V}$ such that $a_{i,n}^\star \in [0,b_i]$, where $k_i= \frac{\gamma_i}{\lambda_i+\mu_{0;i}}$. Substituting back, we get
\begin{equation}\begin{array}{l}
\left[\frac{k_2}{k_1} (x_{0,n}+a_{1,n}^\star)+x_{0,n}+a_{1,n}^\star\right]^2 =k_1 \rho_n(x_{0,n}+a_{1,n}^\star)\\
\Rightarrow  (x_{0,n}+a_{1,n}^\star) \left( k_2 +k_1 \right)^2 = k_1^2 k_2 \rho_n\\
\Rightarrow  a_{1,n}^\star = \left(\frac{ k_1}{k_1 +k_2}\right)^2 k_2\rho_n -x_{0,n}\\
  a_{2,n}^\star = \left(\frac{ k_2}{k_1 +k_2}\right)^2 k_1\rho_n - (1-x_{0,n}).
\end{array} \label{eq:finalNE}
\end{equation}
If the resulting $x_{0,i}^\star \notin (0,b_i)$ from \eqref{eq:finalNE} for some $i$, then we have the other two cases of the proposition. 
\end{IEEEproof}

%\textbf{Special case:} When $B_i \geq N b_i$ is one where we can get a closed form solution for the best response is when $B_i \geq N b_i$ for $i \in \{1,2\}$. This means that the firm does not have a tight budget for the expenditure and each agent can potentially receive the maximum investment of $b_i$ if it is profitable. In this case, we can simply take $\mu_0=0$ while evaluating both of the best responses as the budget constraint will never be violated. A NE of this game is such that $a_{i,n}^* = \beta_{i,n}(x_{0;i},a_{-i}^*)$ for all $i,n$. Thus, we have
%\begin{equation}
%a_{i,n}^*= \min\left\{ b_i, \max \left\{0, \sqrt{\frac{\alpha_{i,n} }{\lambda_i} } - 1 -a_{-i,n}^* \right\} \right\} 
%\end{equation}
%which must be satisfied for all $i,n$. This corresponds to a set of $N$ \emph{independent} fixed point equations which can be easily solved. 
%

\section{Numerical performance analysis}
\label{sec:num}

For all simulations we fix $\gamma_1=\gamma_2=1$, $B_1=B_2=10$ and $\lambda_1=\lambda_2=0.1$ implying symmetry between the two firms in advertising efficiency and profit ratios. 
\subsection{Network aware marketing versus broadcasting}

For a comparison of the proposed graph aware marketing policy with classical policies, we define the \textit{uniform budget allocation (UBA)} policy, \ie a broadcast strategy as 
\begin{equation}
a^{\mathrm{UBA}}_i:=\frac{B_i}{N}.
\end{equation}
This strategy is of particular interest as a broadcast strategy is traditionally used by firms to advertise over media such as television or radio, while social media marketing is targeted to particular users but requires some investment in order to learn the initial opinion of users as well as the network structure.

In Fig.~1, we study the gain a player can make by implementing the best-response strategy \eqref{eq:brf} instead of a UBA strategy with uniform resource allocation. For this simulation, we consider $N=100$ with $\rho_n \in \left\{ 1, C \right\}$. This can be interpreted as the AIP of a collection of disconnected star graphs with the internal nodes being the leaders and $C$ representing the influence the leaders have on the rest. If $n$ is a ``leader", he will have $\rho_n=C$, and $1$ otherwise. We directly consider this configuration for $\rho$ instead of looking at the exact graph structure in order to highlight the message of this numerical example. The resulting difference in utility between the two strategies is referred to as the \emph{gain of targeting} (GoT), and is measured as
\begin{equation}
\text{GoT} := \frac{u_1(x_0,\beta_1(a_2^{\mathrm{UBA}}),a_2^{\mathrm{UBA}})  - u_1(x_0,a_1^{\mathrm{UBA}},a_2^{\mathrm{UBA}})}{u_1(x_0,a_1^{\mathrm{UBA}},a_2^{\mathrm{UBA}})}. \label{eq:gaind}
\end{equation}
 We take $x_n(0)=0.5$ for all $n \in \mathcal{V}$ to remove any bias due to initial opinions. As expected, we observe that a larger disparity in the AIP ($C$) leads to a larger gain by allocating more resources to the leaders. If there are too few leaders, and $C$ is not large enough (as in $C\in\{5,10\}$), this profit saturates resulting in a lower gain. 
\begin{figure}[h]
\begin{center}
\includegraphics[width= 0.49 \textwidth, trim={10cm 0cm 9.5cm 0cm },clip]{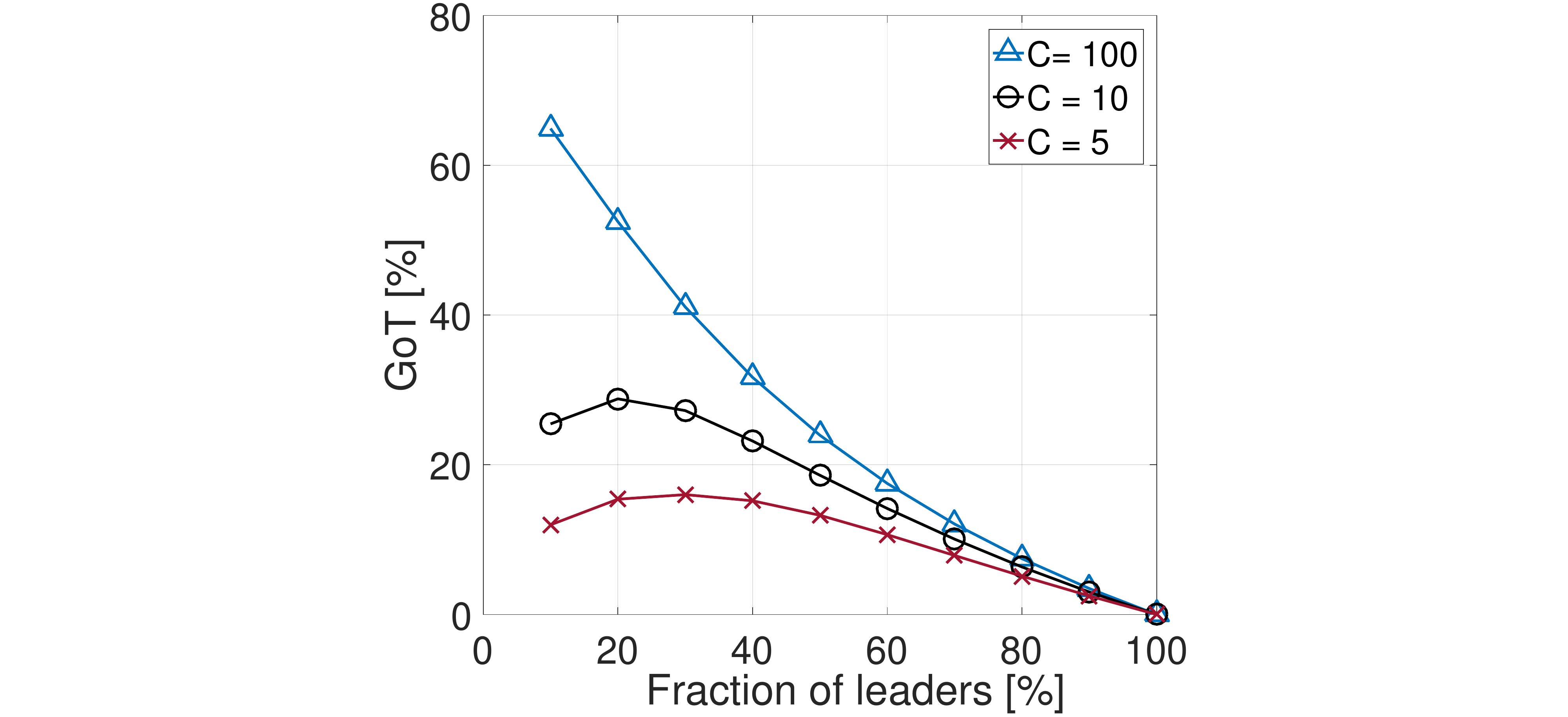}
\caption{Gain by implementing the best response strategy over the uniform budget allocation strategy \eqref{eq:gaind} based on the fraction of leaders.}\label{fig4}
\end{center}\vspace{-0.5cm} \end{figure} 

\subsection{Influence of the initial opinions and graph structure}
\begin{figure}[h]
\begin{center}
\includegraphics[width= 0.26 \textwidth ,trim={0.7cm 4cm 0.7cm 4cm},clip]{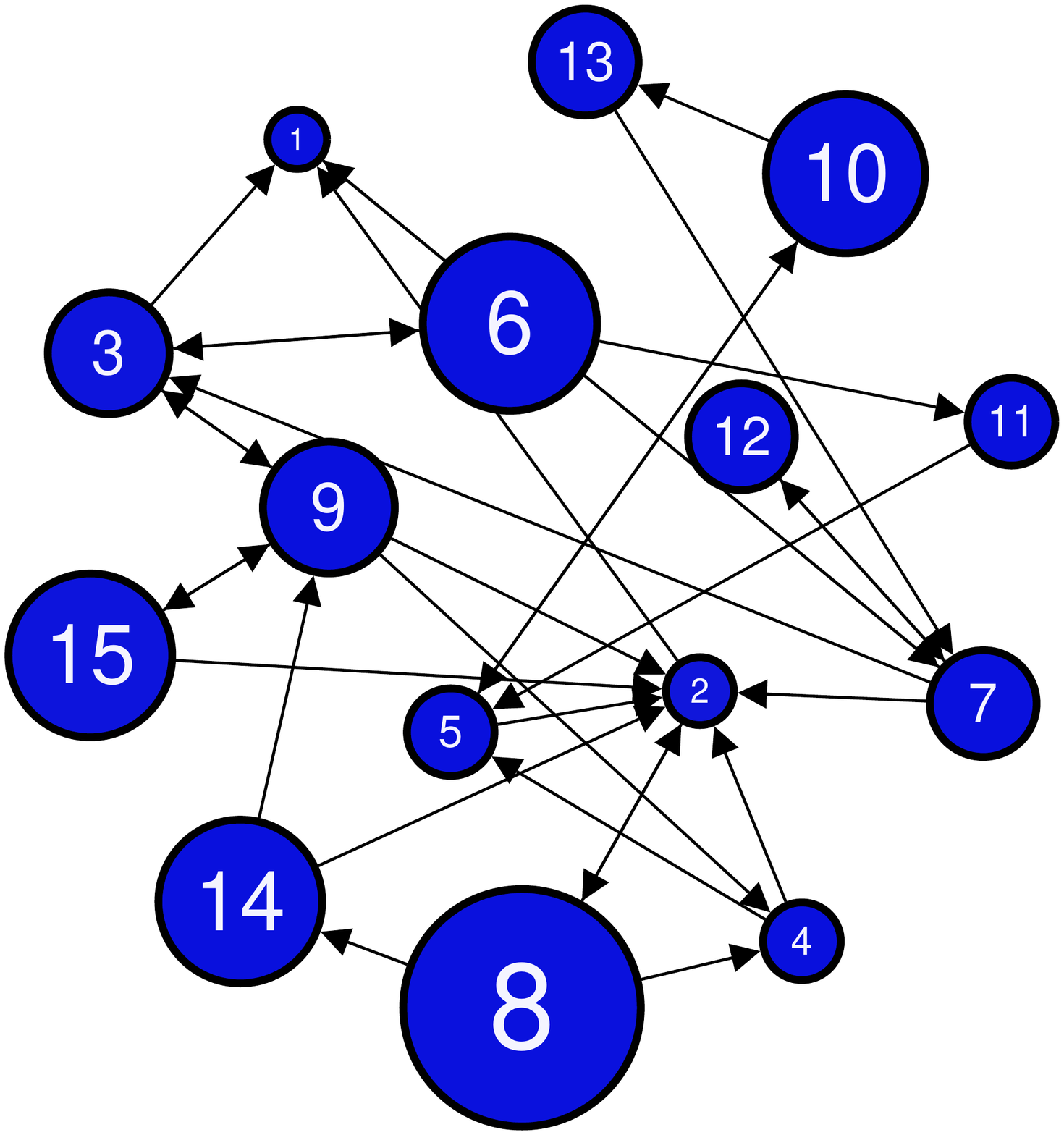}
\caption{The size of the nodes are scaled based on the AIP $\rho_n$.}\label{fig1}
\end{center} \vspace{-.4cm} \end{figure}

For our next numerical analysis, we use the graph structure shown in Fig.~\ref{fig1} with $15$ agents which results in 
\[\begin{array}{r}
\rho = (0.27,0.37,0.98,    0.48,    0.59,   1.58,    0.81,   2.30,   1.09,   1.40\\ ,  0.59 ,   0.81,   0.81  , 1.46,   1.46)
\end{array}\]
by taking the duration $T=10$ and calculating the AIP as $\rho= {1}_N^\top \exp(-10\mathbf{L})$. We initialize the starting opinions to a random opinion given by
\[\begin{array}{r}
x(0) = ( 0.26  ,0.76     ,0.82     ,0.10     ,0.18     ,0.26      ,0.6     ,0.52     ,0.34   \\  ,0.18     ,0.21     ,0.61     ,0.68     ,0.47     ,0.31)^T.
\end{array}\]
\begin{figure}[h]
\centering
 \begin{subfigure}[$N=10$]{\linewidth }
\includegraphics[width= 0.99 \textwidth]{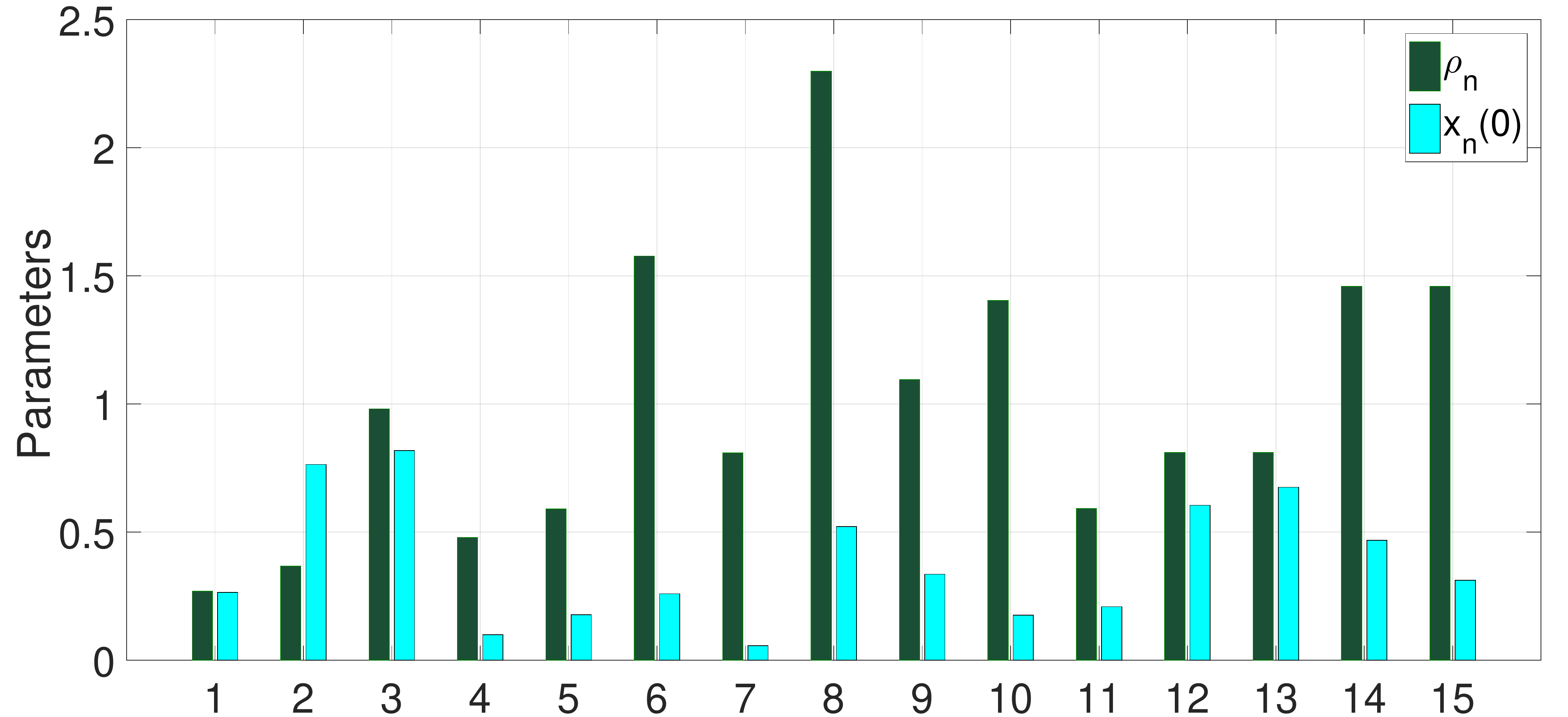} \end{subfigure}
 \begin{subfigure}[$N=10$]{\linewidth }
\includegraphics[width= 0.99 \textwidth]{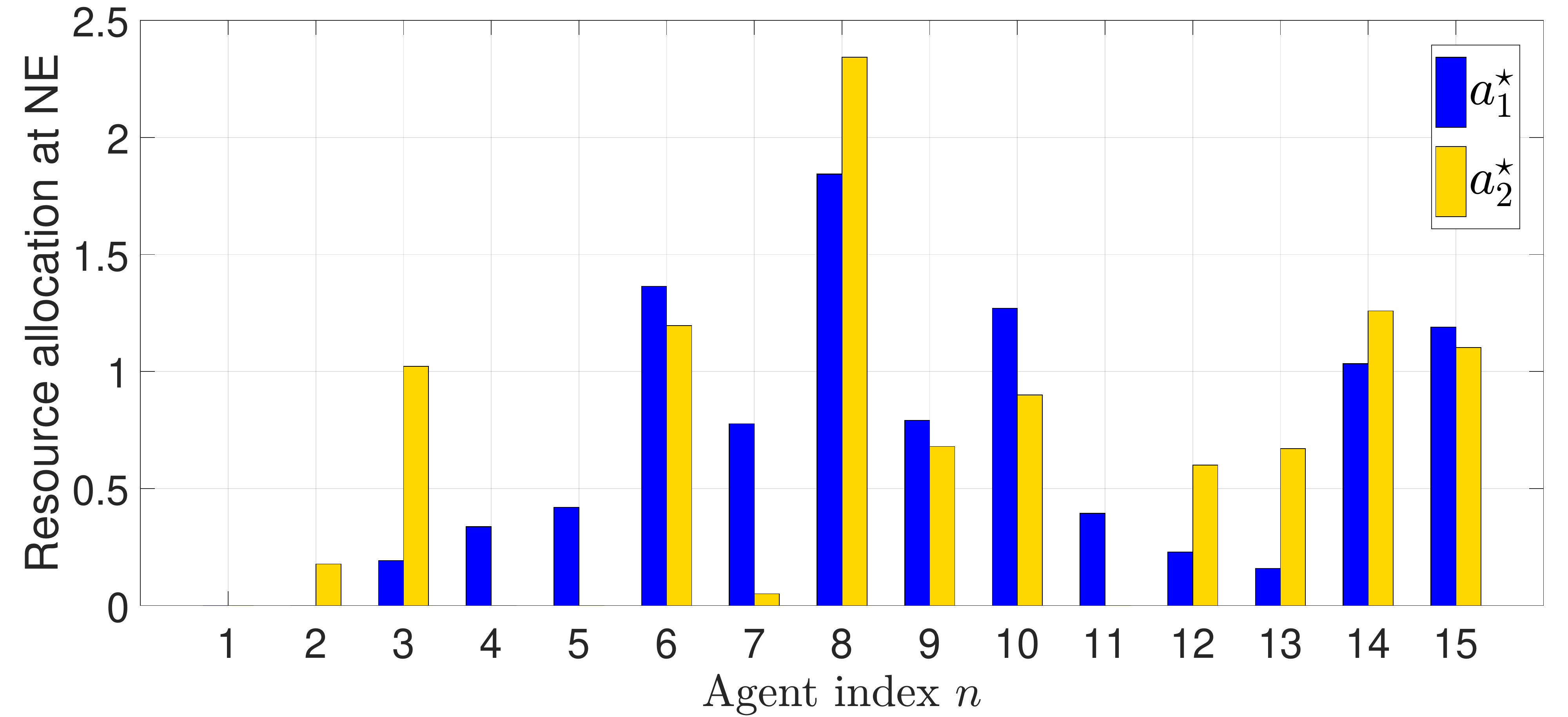} \end{subfigure}
\caption{The sub-figure on top shows the AIP $\rho_n$ and initial opinion $x_n(0)$ for Agent $n$, $n\in\{1,...,N\}$. Corresponding to this initial configuration, the sub-figure on the bottom shows the resource allocation strategies at the NE.}\label{fig3}
\vspace{-.4cm} \end{figure}
 
In Fig.~3, we compare the allocation of budget by the two players at the NE. We observe that if $x_n(0)$ is closer to $1$, \ie  initially biased towards Firm $1$, Firm $2$ will invest more to bring it closer to $0$ while a starting opinion close to $0$ makes Firm $1$ invest more. Both firms invest more on agents with a larger AIP $\rho_n$. This allocation corresponds to the inference from Proposition 2.

\section{Conclusion}
\label{sec:con}

In this paper, we introduce a novel static game model which studies the competition between firms trying to capture a market share by advertising over social media. The consumers of the social network are therefore not only under influence of the other consumers of the network but also of the firms. We conduct a complete equilibrium analysis for the proposed game. Our analysis provides concrete insight into how firms should allocate their budget. In particular, the amount of budget a firm should invest is shown under typical conditions to be proportional to a quantity which we define as the agent influence power (AIP). Interestingly, our analysis allows one to quantify the gain of targeting (GoT) \ie the benefit of implementing a smart allocation policy (based on best-response) instead of allocating the available budget uniformly over the consumers. The corresponding gain is clearly shown to be related to the fraction of dominant consumers and the value of their agent influence power. All these encouraging results show the strong interest in developing the proposed framework. One very relevant extension might be given by a stochastic formulation of the problem.

\bibliography{LCSSfinal}

\begin{thebibliography}{10}
\providecommand{\url}[1]{#1}
\csname url@samestyle\endcsname
\providecommand{\newblock}{\relax}
\providecommand{\bibinfo}[2]{#2}
\providecommand{\BIBentrySTDinterwordspacing}{\spaceskip=0pt\relax}
\providecommand{\BIBentryALTinterwordstretchfactor}{4}
\providecommand{\BIBentryALTinterwordspacing}{\spaceskip=\fontdimen2\font plus
\BIBentryALTinterwordstretchfactor\fontdimen3\font minus
  \fontdimen4\font\relax}
\providecommand{\BIBforeignlanguage}[2]{{%
\expandafter\ifx\csname l@#1\endcsname\relax
\typeout{** WARNING: IEEEtran.bst: No hyphenation pattern has been}%
\typeout{** loaded for the language `#1'. Using the pattern for}%
\typeout{** the default language instead.}%
\else
\language=\csname l@#1\endcsname
\fi
#2}}
\providecommand{\BIBdecl}{\relax}
\BIBdecl

\bibitem{singh1984price}
N.~Singh and X.~Vives, ``Price and quantity competition in a differentiated
  duopoly,'' \emph{The RAND Journal of Economics}, pp. 546--554, 1984.

\bibitem{friedman1958game}
L.~Friedman, ``Game-theory models in the allocation of advertising
  expenditures,'' \emph{Operations Research}, vol.~6, no.~5, pp. 699--709,
  1958.

\bibitem{esmaeili2009game}
M.~Esmaeili, M.-B. Aryanezhad, and P.~Zeephongsekul, ``A game theory approach
  in seller--buyer supply chain,'' \emph{European Journal of Operational
  Research}, vol. 195, no.~2, pp. 442--448, 2009.

\bibitem{DeGroot}
M.~H. DeGroot, ``Reaching a consensus,'' \emph{Journal of the American
  Statistical Association}, vol.~69, no. 345, pp. 118--121, 1974.

\bibitem{tsang2004consumer}
M.~Tsang, S.-C. Ho, and T.-P. Liang, ``Consumer attitudes toward mobile
  advertising: An empirical study,'' \emph{International journal of electronic
  commerce}, vol.~8, no.~3, pp. 65--78, 2004.

\bibitem{woerndl2008internet}
M.~Woerndl, S.~Papagiannidis, M.~Bourlakis, and F.~Li, ``Internet-induced
  marketing techniques: Critical factors in viral marketing campaigns,''
  \emph{International Journal of Business Science and Applied Management},
  vol.~3, no.~1, 2008.

\bibitem{Camponigro}
M.~Caponigro, B.~Piccoli, F.~Rossi, and E.~Tr\'elat, ``Sparse feedback
  stabilization of multi-agent dynamics.'' in \emph{Proceedings of the 55th
  IEEE Conference on Decision and Control}, 2016, pp. 4278--4283.

\bibitem{Dietrich}
F.~Dietrich, S.~Martin, and M.~Jungers, ``Control via leadership of opinion
  dynamics with state and time-dependent interactions,'' \emph{IEEE Trans. on
  Automatic Control}, vol. 10.1109/TAC.2017.2742139, 2017.

\bibitem{Hegselmann}
R.~Hegselmann, S.~Kurz, C.~Niemann, and J.~Rambau, ``Optimal opinion control :
  The campaign problem,'' \emph{Journal of Artificial Societies and Social
  Simulation}, vol.~18, no.~3, 2015.

\bibitem{domingos2001mining}
P.~Domingos and M.~Richardson, ``Mining the network value of customers,'' in
  \emph{Proceedings of the seventh ACM SIGKDD international conference on
  Knowledge discovery and data mining}.\hskip 1em plus 0.5em minus 0.4em\relax
  ACM, 2001, pp. 57--66.

\bibitem{arthur2009pricing}
D.~Arthur, R.~Motwani, A.~Sharma, and Y.~Xu, ``Pricing strategies for viral
  marketing on social networks,'' in \emph{International Workshop on Internet
  and Network Economics}.\hskip 1em plus 0.5em minus 0.4em\relax Springer,
  2009, pp. 101--112.

\bibitem{lasaulce-book-2011}
S.~Lasaulce and H.~Tembine, \emph{Game Theory and Learning for Wireless
  Networks~: Fundamentals and Applications}.\hskip 1em plus 0.5em minus
  0.4em\relax Academic Press, 2011.

\bibitem{masucci2014strategic}
A.~M. Masucci and A.~Silva, ``Strategic resource allocation for competitive
  influence in social networks,'' in \emph{Communication, Control, and
  Computing (Allerton), 2014 52nd Annual Allerton Conference on}.\hskip 1em
  plus 0.5em minus 0.4em\relax IEEE, 2014, pp. 951--958.

\bibitem{varma2017opinion}
V.~Varma, I.-C. Morarescu, S.~Lasaulce, and S.~Martin, ``Opinion dynamics aware
  marketing strategies in duopolies,'' in \emph{56th IEEE Conference on
  Decision and Control, CDC 2017}, 2017.

\bibitem{martins2008continuous}
A.~Martins, ``Continuous opinions and discrete actions in opinion dynamics
  problems,'' \emph{International Journal of Modern Physics C}, vol.~19,
  no.~04, pp. 617--624, 2008.

\bibitem{nash1951non}
J.~Nash, ``Non-cooperative games,'' \emph{Annals of Mathematics}, pp. 286--295,
  1951.

\bibitem{rosen1965existence}
J.~B. Rosen, ``Existence and uniqueness of equilibrium points for concave
  n-person games,'' \emph{Econometrica: Journal of the Econometric Society},
  pp. 520--534, 1965.

\bibitem{gilboa1991social}
I.~Gilboa and A.~Matsui, ``Social stability and equilibrium,''
  \emph{Econometrica: Journal of the Econometric Society}, pp. 859--867, 1991.

\bibitem{matsui1992best}
A.~Matsui, ``Best response dynamics and socially stable strategies,''
  \emph{Journal of Economic Theory}, vol.~57, no.~2, pp. 343--362, 1992.

\bibitem{hofbauer2006best}
J.~Hofbauer and S.~Sorin, ``Best response dynamics for continuous zero-sum
  games,'' \emph{Discrete and Continuous Dynamical Systems Series B}, vol.~6,
  no.~1, p. 215, 2006.

\bibitem{boyd2004convex}
S.~Boyd and L.~Vandenberghe, \emph{Convex optimization}.\hskip 1em plus 0.5em
  minus 0.4em\relax Cambridge University Press, 2004.

\end{thebibliography}

\end{document}